\documentstyle[sprocl]{article}
\def\Journal#1#2#3#4{{#1} {\bf #2}, #3 (#4)}
\def\PRD{{\em Phys. Rev.} D}

\def\be{\begin{equation}}
\def\ee{\end{equation}}
\def\bea{\begin{eqnarray}}
\def\eea{\end{eqnarray}}
\def\al{\alpha}
\def\bt{\beta}

\def\Ga{\Gamma}
\def\de{\delta}

\def\la{\lambda}
\def\La{\Lambda}
\def\vr{\varrho}
\def\si{\sigma}

\def\pt{\partial}
\def\na{\nabla}

\def\G{{\Gamma}}

\def\al{\alpha}
\def\bt{\beta}

\def\Ga{\G}
\def\de{\delta}

\def\la{\lambda}
\def\La{\Lambda}
\def\vr{\varrho}
\def\si{\sigma}

\def\pt{\partial}
\def\na{\nabla}

\def\G{\Gamma}

\begin{document}

\title{ALTERNATIVE LAGRANGIANS FOR EINSTEIN METRICS
\footnote{Proceedings of the International Seminar on 
Mathematical Cosmology, Potsdam, March 30--April 4, 1998, M. Rainer
and H.-J. Schmidt (eds.), World Scientific PC Singapore.}}

\author{A. BOROWIEC}

\address{Institute of Theoretical Physics, Wroc{\l}aw University, 
pl. Maxa Borna 9,\\ 50-204 Wroc{\l}aw, Poland\\E-mail: borow@ift.uni.wroc.pl} 

\author{M. FRANCAVIGLIA}

\address{Departimento di Matematica, Universit\`a di Torino, 
Via C. Alberto 10,\\
10123 Torino, Italy\\E-mail: fviglia@dm.unito.it}
\maketitle
\abstracts{We shall use the variational decomposition
technique in order to calculate equations of motion and Noether 
energy-momentum complex for some classes of non-linear gravitational 
Lagrangians within the first-order (Palatini) formalism.
In particular, a complex space-time appears as a solution of 
our variational problem.}

\section{Introduction}
We report on recent results  stating that some classes of 
non-linear gravitational Lagrangians give, in the first-order
formalism, Einstein field equations and the Komar expression
for the energy-momentum complex. Such Lagrangians are particular 
important since, at the classical level, they are equivalent to
General Relativity. However, their quantum contents and 
divergences could be slightly improved.
This note is based on joint works with 
M. Ferraris (Torino) and I. Volovich (Moscow).

\subsection{Variational Decomposition and Noether Theorems}

It is well know that a variation (i.e. functional derivative) of an
arbitrary-order Lagrangian  
$L(\phi)\equiv L(\phi,\phi_\mu,\phi_{\mu\nu},\ldots)$ 
 \footnote{
For simplicity, we drop an internal field index, e.g. $\phi^A$.}
decomposes into two parts according to the "first variation formula" 
\footnote{
We adopt the Einstein summation convention.}
\be
\de L= \frac{\de L}{\de\phi} \de\phi + \pt_\mu \vr^\mu 
\ee
Here $\phi_\mu=\pt_\mu\phi, \ldots$ denotes the partial derivatives 
of $\phi$ with respect to (local) space-time (independent) variables
$x^\mu$, $\mu=1,\ldots,n$.
The first term represents the Euler-Lagrange expression,
i.e. field equations. The second part is a divergence of 
$\vr^\mu\equiv\vr^\mu (\phi, \de\phi)$, where
$$
\vr^\mu=[\frac{\pt L}{\pt\phi_\mu}-\pt_\nu (\frac{\pt L}
{\pt\phi_{\mu\nu}})]\de\phi + \frac{\pt L}{\pt\phi_{\mu\nu}}\de\phi_\nu
+\ldots $$
Although this second (boundary) term does not contribute to the equations
of motion it is physically important since it does contribute to 
the conservation laws (Noether Theorems).

For the variation $\de_*\phi$ implemented by an (infinitesimal) symmetry 
transformation  one has $\de_* L=\pt_\mu\tau^\mu$
without using  the equations of motion. Therefore,
equation (1) can be rewritten under the following form
$$
\frac{\de L}{\de\phi} \de_*\phi = - \pt_\mu (\vr_*^\mu -\tau^\mu) 
$$
where $\vr_*^\mu = \vr^\mu (\phi, \de_*\phi)$. A {\it Noether current}
then arises
\be
E^\mu\equiv E^\mu (\phi, \de_*\phi) = \vr_*^\mu -\tau^\mu \ee
which is conserved {\it on shell}, i.e. when the field equations are 
satisfied. One writes $\pt_\mu E^\mu\approx 0$
and calls it a {\it weak conservation law}. 
In the present paper we deal with so-called 
{\it local symmetries}  (and second Noether's Theorem). 
In this case, there exists 
a skew-symmetric quantity  $U^{\mu\nu}=-U^{\nu\mu}$, called a
{\it superpotential} (see e.g.\cite{FF3,BFF}), such that
$$E^\mu\approx\pt_\nu U^{\mu\nu}$$
i.e. $E^\mu$ differs from the divergence $\pt_\nu U^{\mu\nu}$ by 
a quantity which vanishes on shell. 
\subsection{Second-Order Einstein-Hilbert Lagrangian}
Einstein metrics are extremals of the Einstein-Hilbert purely
metric variational problem.
Consider the Einstein-Hilbert (linear) gravitational Lagrangian 
\be
L_{H}(g, \pt g, \pt^2 g) = 
%|det g|^{\frac{1}{2}} (g^{\mu\nu} R_{\mu\nu} +(2-n)\La)
|det g|^{\frac{1}{2}} (R - c)
\ee
Here standard notation for the Riemann and Ricci tensor 
\bea
R^\al_{\bt\mu\nu} & = & R^\al_{\bt\mu\nu}(g) =
\pt_\mu\Ga^\al_{\bt\nu}-\pt_\nu\Ga^\al_{\bt\mu} +
\Ga^\al_{\si\mu}\Ga^\si_{\bt\nu}-\Ga^\al_{\si\nu}\Ga^\si_{\bt\mu}
\nonumber\\
R_{\mu\nu} & = & R_{\mu\nu}(g)=R^\al_{\mu\al\nu}
\eea
of the Levi-Civita connection on a space-time manifold $M$ 
($\hbox{\rm dim} M=n$)
\be
\Ga^\al_{\bt\mu}(g)= \frac{1}{2}g^{\al\si}(\pt_\bt g_{\mu\si} +\pt_\mu
g_{\si\bt} - \pt_\si g_{\bt\mu}) \ee
is in use.
In the Lagrangian above $R=R(g)=g^{\mu\nu}R_{\mu\nu}(\Ga)$ denotes 
the scalar curvature. 
In this way the metric $g$ becomes the only dynamical variable of 
the theory. According to well known formula
$\de\sqrt g =-\frac{1}{2}\sqrt gg_{\al\bt}\de g^{\al\bt}$,
variation of $L_{H}$ with respect of an arbitrary variation of $g$ reads
\footnote{
We simply write $\sqrt g$ instead of $\sqrt{|det g|}$.}
$$
\de L_{H}=\sqrt g [R_{\al\bt}-1/2 (R-c)g_{\al\bt}]
\de g^{\al\bt}+ \sqrt g g^{\al\bt} \de R_{\al\bt}
$$
Taking into account that from (4)
\be
\de R_{\al\bt}=\na_\mu\de\Ga^\mu_{\al\bt}- \na_\al\de\Ga^\si_{\bt\si}
\ee
and that covariant derivatives $\na_\al$ for the Levi-Civita connection
of $g$ commutes with $\sqrt g g^{\al\bt}$, we get
\be
\sqrt gg^{\al\bt}\de R_{\al\bt}=\na_\mu [\sqrt g
g^{\al\bt}(\de\Ga^\mu_{\al\bt}-\de^\mu_\al\de\Ga^\si_{\bt\si})]
\ee
Quantity in the square brackets transforms as a vector density of
weight $1$.
It allows to replace the covariant derivatives $\na_\mu$ in (7)
by the partial one $\pt_\mu$.
\footnote{Since one deals with a symmetric connection.} 

Therefore, a variational decomposition for $L_{H}$ takes finally the form
\be
\de L_{H}=\sqrt g [R_{\al\bt}-1/2 (R-c)g_{\al\bt}]
\de g^{\al\bt}+\pt_\mu [\sqrt g g^{\al\bt}(\de\G^\mu_{\al\bt}-
\de^\mu_\bt\de\G^\si_{\al\si})]
\ee
This produces, of course, the Einstein field equations for the metric $g$ 
\be
R_{\mu\nu}(g)=\La g_{\mu\nu}\ee
with the cosmological constant $\La=c/(n-2)$.\footnote{
In this letter we always assume $n>2$.}

As a symmetry transformation, consider now a $1$-parameter group of
diffeomorphisms generated by the vectorfield $\xi=\xi^\al\pt_\al$ on 
$M$. In this case one  can utilize the well known expressions
\bea
\de_* g & = & \de_\xi g\equiv {\cal L}_\xi g_{\al\rho}=
\na_{\al}\xi_{\rho} + \na_{\rho}\xi_{\al} \nonumber\\
\de_*\Ga & = & \de_\xi\Ga^\bt_{\al\rho}\equiv {\cal L}_\xi 
\Ga^\bt_{\al\rho}=\xi^\si R^\bt_{\al\si\rho} +\na_{\al}\na_{\rho}\xi^\bt
\eea
where ${\cal L}_\xi$ stands for the Lie derivative along $\xi$.
Our Lagrangian is {\it reparametrization invariant}, 
in the sense that diffeomorphisms of $M$ transform $L_{H}$ 
as a scalar density of weight $1$.  This
means that, at the infinitesimal level, one has
\be
\de_*L_H={\cal L}_{\xi}L_H=\pt_\al(\xi^\al L_H) 
\ee
As a consequence, equation (2), in this case, can be written as follows:
\be
E^\mu(\xi)=\sqrt g(g^{\al\bt}\de^\mu_\si-g^{\al\mu}\de^\bt_\si)
\na_{(\al}\na_{\bt )}\xi^\si + (2\sqrt g R^\mu_\si-\de^\mu_\si L_{H})\xi^\si 
\ee
This provides the global and covariant expression for the Noether
energy-momen\-tum flow of a gravitational field represented by the 
Einstein metric $g$ and calculated along a vectorfield $\xi$.
The corresponding superpotential \cite{Kij,FF3}
\be
U_{H}^{\mu\nu}(\xi) = |det g|^{\frac{1}{2}}
(\na^\mu\xi^\nu - \na^\nu\xi^\mu)
\ee
is known as the {\it Komar superpotential}.\cite{Go,Katz}
Problems with the definition of gravitational energy and momentum
appear when one tries to make (12-13) independent of the
vectorfield $\xi$.\cite{Kij,FF3,Gib} An interesting application
of the Komar expression to the black hole entropy has been
presented in \cite{IW}.

\section{Non-Linear First-Order Lagrangians}
It is known that the non-linear Hilbert type Lagrangians $f(R)\sqrt g$,
where $f$ is a function of one real variable, lead to fourth order equation
for $g$, which are not equivalent to Einstein equations unless $f(R)=R-c$
(linear case), or to appearance of additional matter fields. 
It is also known that the linear "first order" Lagrangian
$r\sqrt g$, where $r=r(g,\Ga)=g^{\al\bt}r_{\al\bt}(\Ga)$ is a scalar 
concomitant of the metric $g$ and  linear (symmetric) connection $\Ga$,
leads to separate  equations for $g$ and $\Ga$ which turn out to be 
equivalent to Einstein equations for $g$.
In the sequel we shall use small letters $r^\al_{\bt\mu\nu}$ and
$r_{\bt\nu}=r^\al_{\bt\al\nu}$ to denote the Riemann and Ricci tensor
of an arbitrary (symmetric) connection $\Ga$ (still given by the same
formulae (4)), i.e. without assuming that $\Ga$ is the Levi-Civita
connection of $g$.

\subsection{Hilbert Type Lagrangians}

As we explained above inequivalence with General Relativity could 
also hold for non-linear first-order Lagrangians
\footnote{ 
Such Lagrangians have been investigated in \cite{FFV}.} 
\be
L_f(g,\Ga)=\sqrt g f(r)
\ee
Now, the scalar
$r(g,\Ga)=g^{\al\bt}r_{\al\bt}(\Ga)$ is not longer the scalar curvature, 
since $\Ga$ is not longer Levi-Civita connection of the metric $g$. We 
choose a metric and a symmetric connection as independent dynamical 
variables (so-called Palatini method, see also \cite{HK}).
Variation of $L_f$ gives
$$
\de L_f=\sqrt g (f^\prime (r)r_{\al\bt}-1/2f(r)g_{\al\bt})
\de g^{\al\bt}+
\sqrt gf^\prime (r)g^{\al\si}\de r_{\al\bt}$$
Substituting $\de r_{\al\bt}$ by an analog of (6) with $\na_\al$ being 
the covariant derivative with respect to $\Ga$ and applying the 
covariant Leibniz rule
("integrating by parts") give rise to the variational decomposition 
\bea
\de L_f & = & \sqrt g (f^\prime (r)r_{\al\bt}-1/2f(r)g_{\al\bt})\de 
g^{\al\bt}-\na_\bt[\sqrt g f^\prime (r)(g^{\al\si}\de^\bt_\la 
\nonumber \\
&&{} - g^{\al\bt}\de^\si_\la)]\de\G^\la_{\al\si} +
\pt_\mu [\sqrt gf^\prime (r) g^{\al\bt}(\de\G^\mu_{\al\bt}-
\de^\mu_\bt\de\G^\si_{\al\si})]
\eea
First observe that the boundary term in (15) apart of the factor
$f^\prime (r)$ is exactly the same as in the Einstein-Hilbert case (8).
Field equations in this case are \cite{FFV}:
\be
f^\prime (r)r_{(\mu\nu)} - \frac{1}{2}f(r)g_{\mu\nu}=0
\ee
\be
\na_\al [f^\prime (r)\sqrt gg^{\mu\nu}] = 0
\ee
where $()$ denotes symmetrization. 
In fact, variation of $L_f$ with respect to $\Ga$ leads to
the following equations (see also (15)):
$$
\na_\bt[\sqrt g f^\prime (r)(g^{\al\si}\de^\bt_\la 
- g^{\bt (\al}\de^{\si )}_\la)]=0
$$
which due to the symmetry of $g^{\mu\nu}$ reduce to (17).
Notice that (16) are not yet
Einstein equations, even when  $f(r)=r$.  
Equations (16-17) must be considered together with the consistency
condition obtained by contraction of (16) with $g^{\mu\nu}$.
It gives then 
\be
f^\prime (r)r-\frac{n}{2}f(r)=0\ee
This equation (except the case it is identically satisfied) 
forces $r$ to take a  set of constant values 
$r=c$, with $c$ being solution of (18). In the generic case 
(simple roots, with
$f^\prime (c)\neq 0$, $n>2$) equation (17) gives 
$$
\na_\al(\sqrt gg^{\mu\nu})=0
$$
which, in turn,  forces $\Ga$ to be the Levi-Civita connection 
of $g$. Replacing back into (16) we find
$$
R_{\mu\nu}(g)=\La (c)g_{\mu\nu}
$$
Einstein equations for the metric $g$ with $\La (c)=f(c)/2f^\prime 
(c)=c/n$. As we  observed above, the boundary term in 
(15) is proportional with the factor 
$f^\prime (c)$  to that of (8). Therefore, energy-momentum 
flow as well as superpotential are proportional to already 
known from the standard Einstein-Hilbert formalism (12-13).
\cite{BFFV,GS} It shows {\it universality} of Einstein equations 
and Komar 
superpotential, i.e. their independence on the choice of the Lagrangian 
(represented by the function $f$). These properties hold true in any 
dimension $n>2$.\footnote{
See \cite{FFV3} for $n=2$ case where non-generic cases have been
also consiered.}

\subsection{Ricci Squared Lagrangians} 
As the next examples consider the family of non-linear
gravitational Lagrangians
\be
\hat L_f(g,\Ga)=\sqrt g f(s)
\ee
parameterized by the real function $f$ of one variable.\cite{BFFV2} 
Now, the scalar ({\it Ricci squared}) concomitant  
$s=s(g,\Ga)=g^{\al\mu}g^{\bt\nu}s_{\al\bt}s_{\mu\nu}$, where 
$s_{\mu\nu}=r_{(\mu\nu)}(\Ga)$ is the symmetric part of the Ricci 
tensor of $\Ga$. Variational decomposition formula reads 
\bea
\de\hat L_f & = & \sqrt g(2f^\prime (s)g^{\mu\nu}s_{\al\mu}s_{\bt\nu}-
\frac{1}{2}f(s)g_{\al\bt})\de g^{\al\bt}- 
\na_\nu [2\sqrt g f^\prime (s)(s^{\al\bt}\de^\nu_\la 
\nonumber \\
&&{} - s^{\nu\al}\de^\bt_\la)]\de\Ga^\la_{\al\bt}+
\pt_\mu [2\sqrt gf^\prime (s)g^{\al\nu}g^{\bt\la} s_{\nu\la}
(\de\G^\mu_{\al\bt}-\de^\mu_\bt\de\G^\si_{\al\si})] 
\eea
where for short $s^{\al\bt}=g^{\al\mu}g^{\bt\nu}s_{\mu\nu}$.
Observe again that an essential part of the boundary term in (20)
coincides with  the previous cases (8,15).
Euler-Lagrange field equations are
\be
f^\prime (s)g^{\mu\nu}s_{\al\mu}s_{\bt\nu}-\frac{1}{4}f(s)g_{\al\bt}=0
\ee
\be
\na_\la (\sqrt gf^\prime (s)g^{\al\mu}g^{\bt\nu}s_{\mu\nu})=0
\ee
Contraction of (21) with $g^{\al\bt}$ gives the consistency
equation
\be
f^\prime (s)s-\frac{n}{4}f(s)=0
\ee
Restricting our attention again to the generic case we find that
for regular solutions $s=c\neq 0$ of (23) ($f^\prime (c)\neq 0$, $n>2$) 
equation (21) can be rewritten in the following (matrix) form \cite{BFFV2}
\be
(g^{-1} h)^2 =\frac{c}{|c|} I
\ee
where $c/|c|=\pm 1$ and
\be
h_{\al\bt}= \sqrt{\frac{n}{|c|}}\ s_{\al\bt}(\Ga) .
\ee
$h_{\al\bt}$ is a symmetric, twice-covariant and due to (21) 
non-degenerate tensor field on $M$ i.e., it is simply a metric.
By making use of the Ans\"{a}tz (25), equations (22) can be converted 
into the form
$$
\na_\la (\sqrt h h^{\al\bt})=0
$$
with $h^{\al\bt}$ being the inverse of $h_{\al\bt}$.
Therefore, the connection $\Ga$  has to be a Levi-Civita
connection for the metric $h$ and as a consequence, (25) becomes 
an Einstein equation for $h$ with the cosmological constant 
$\La=\sqrt{|c|/n}$. 
Substituting further (25) into the boundary term in (20) we find that, 
up to a constant multiplier,
the energy-momentum flow and the superpotential are given by the same 
expressions as (12-13) with the metric $g$ replaced by $h$.
This extends a notion of universality also to the class of Ricci 
squared Lagrangians.\cite{BFFV2}

The algebraic constraints (24) are of special interest by their own.
They provide on space-time some additional 
differential-geometric structures, namely
a Riemannian almost-product structure and/or an 
almost-complex anti-\linebreak[4]
Hermitian ($\equiv$ Norden) structure.\cite{BFFV3}

In the (psedo-)Riemannian almost-product case one equivalently deals 
with an almost-product structure given by the $(1,1)$ tensor field 
$P=g^{-1}h$ ($P^2=I$) as well as with a compatible metric $h$
satisying the condition \footnote{
In our case the metric $h$ should be in addition Einsteinian.}
\be
h(PX, PY) = h(X, Y)
\ee
which is also encoded in the simple algebraic relation (24).
Here $X, Y$ denote two arbitrary vecorfields on $M$.

There is a wide class of integrable
almost-product structures, namely so called {\it warped product} 
structures \cite{Besse,CC}, which are an intrinsic property of some 
well know exact solutions of Einstein equations: these include 
e.g. Schwarzschild, Robertson-Walker, Reissner-Nordstr\"{o}m, 
de Sitter, etc. (but not Kerr!).
Some other examples are provided by Kaluza-Klein type theories,
$3+1$ decompositions and more generally so called 
{\it split} structures \cite{GK}. The explicite form of the zeta
function on product spaces and of the multiplicative anomaly
has been derived recently in \cite{BW}.

In the anti-Hermitian case one deals with $2m$ - dimensional 
manifold $M$, an almost complex structure $J=g^{-1}h$ ($J^2=-I$)
and an anti-Hermitian \footnote{Recall that for 
a Hermitian metric $h(JX, JY) = h(X, Y)$.} metric $h$:
\be
h(JX, JY) = - h(X, Y)\ee
This implies that the signature of $h$ should be $(m,m)$.
In the K\"ahlerian case
($\na J = 0$ for the Levi-Civita connection of $h$) the almost-complex
structure is automaticly integrable. We have proved \cite{BFFV3} that 
in fact the metric $h$ has to  be a real part of certain  holomorphic 
metric on a complex (space-time) manifold $M$. 

It should be however remarked that a theory of complex manifolds
with holomorphic metric (so called {\it complex Riemannian} manifods) has
become one of the corner-stone of the twistor theory \cite{Fl}.
This includes a {\it non-linear graviton} \cite{Pe}, 
theory of ${\cal H}${\it -spaces} \cite{BFP} and {\it ambitwistor} 
formalism \cite{LeB}.
\subsection{Conclusions}
We showed that the use of Palatini formalism leads to results 
essentially different from the metric formulation when one deals with 
non-linear Lagrangians: with the exception of special ("non-generic") 
cases we always obtain the Einstein equations  as gravitational field 
equations and Komar complex as a Noether energy-momentum complex. In 
this sense non-linear (matter-free)\footnote{
See e.g. \cite{Rub} how to include matter.} 
theories are equivalent to General Relativity: 
they admit two families of alternative Lagrangians (14, 19) for 
the Einstein equations with a cosmological constant. 
In $n=2$ dimensions, they provide a general mechanism for
governing topology change.\cite{FFV3}

Moreover, in the case of Ricci squared Lagrangians (19), 
besides the initial metric $g$ one gets the Einstein metric $h$. 
Both metrics are related by algebraic equation (24).
These aspects have been considered in \cite{BFFV2}. 
A characterization and examples of anti-K\"ahler Einstein manifolds 
as well as almost-product Einstein manifolds has been 
obtained. \cite{BFFV3}

Our results can be relevant for quantum gravity.
In fact, in order to remove divergences  one has to add counterterms
to the Lagrangian which depend not only on the scalar curvature but also
on the Ricci and Riemann tensor invariants. It follows from our results
that in the first order formalism, such counterterms do not change the
semiclassical limit, since genericly we still have the standard Einstein
equation. 

\section*{Acknowledgments}
One of us (A.B.) acknowledges a grant from the 
{\em Deutsche Physikalische Gesells\-chaft} and
{\em Polskiego Towarzystwa Fizycznego}. 
%for a collaboration which made his visit to Potsdam possible.

\end{document}